\documentclass[pre,twocolumn,showpacs,showkeys,amsmath]{revtex4}
\usepackage{graphicx}
\begin{document}

\title{Mean-Field and Non-Mean-Field Behaviors in Scale-free Networks with Random Boolean Dynamics}

\author{A. Castro e Silva$^{1,\ast}$ and J. Kamphorst Leal da Silva$^{2,\dagger}$}

\affiliation{
$^{1}$ Departamento de F\'\i sica,
Universidade Federal de Ouro Preto\\
Campus Universit\'ario, 35.400-000 Ouro Preto/MG, Brazil\\
$^{2}$ Departamento de F\'\i sica,
Universidade Federal de Minas Gerais\\
Caixa Postal 702, 30.161-970, Belo Horizonte/MG, Brazil\\}

\begin{abstract}
We study two types of simplified Boolean dynamics in scale-free networks, both with synchronous update. Assigning only Boolean functions AND and XOR to the nodes with probability $1-p$ and $p$, respectively, we are able to analyze the density of $1$'s and the Hamming distance on the network by numerical simulations and
by a mean-field approximation (annealed approximation). We show that the behavior is quite different if the node always enters in the dynamics as its own input (self-regulation) or not. The same conclusion holds for the Kauffman NK model. Moreover, the simulation results and  the mean-field ones (i) agree well when there is no self-regulation, and (ii) disagree for small $p$ when self-regulation is present in the model.  

\end{abstract}

\maketitle

\section{Introduction}

Some physicists have claimed that it is possible to roughly classify its science branches in physics of  small,  big and finally complex systems. Although such classification may appear simple, it suits very well for the present work. Complex systems are known as entities composed of a large number of agents sharing  a rich set of simple and non linear interactions. In such systems, different behaviors can be achieved if the interaction change, even if the interacting agent remains the same. These systems are well modeled by using network concepts, where the agents are called nodes, and interactions appear as links. 

Since the release of the Barab\'{a}si and Albert paper   \cite{barabasi1999emergence} about
growing networks, a lot of knowledge has been achieved regarding the topology and 
properties of such objects   \cite{barabási2004network,dorogovtsev2002condensed}. In addition, it was found 
that these types of networks can be found in a variety of fields like social relation, voting, disease spreading, the WWW, neural and regulatory networks  \cite{wardil2008discrete,bernardes2002epidemic,watts1998small,redner1998popular,
huberman1998strong,albert1999diameter,adamic1999growth,
barthélémy1999small,newman1999renormalization,barrat2000properties,
amaral2000classes,barbosa2006scaling}.

The interaction among nodes of a network can be modeled in very different ways. In the 
most simple models, it is assumed that (i) nodes can only have two states $0$ (\textit{off}) or $1$ (\textit{on}), and that (b) the dynamics is done via Boolean functions. These models, very suitable for  simulations \textit{in silico}, can be be very useful to study agents that interact using \textit{on/off} states as happens in gene expression/repression   \cite{jeong2000large} and  protein activation/inhibition \cite{barabási2004network}. Models involving Boolean dynamics are called Random Boolean Networks ($RBNs$). The first $RBN$ model was introduced by Stuart Kauffman, and it is known as $NK$ model since it was composed by $N$ nodes,  
each one with its own random $K$ inputs, and with the Boolean functions randomly chosen \cite{kauffman1969metabolic}.

In general, all $RBNs$ share some common features. They are directed networks with $N$ nodes, and a node $i$ has $k_i$ inputs that regulate its next state via some random Boolean  function $G_i$. The topology of the interactions can be expressed by using a proper connectivity distribution of inputs $P(k)$ and the 
type of the Boolean functions can  be changed in order  to adequate the restrictions imposed by the problem.  The Boolean functions can be chosen basically in two different ways: (a) all functions are chosen \textit{a priori}, and they are kept 
constant during time evolution - the so-called \textit{quenched} models; and (b) the functions change during the dynamics, meaning that
for each time step, each node have a new function - the so-called \textit{annealed} models. Another important 
dynamical feature  is the update of the nodes. In the synchronous or parallel 
mode, all nodes are updated simultaneously in each time step. On the other hand, in the asynchronous or serial mode,   first we randomly choose a node that is updated immediately; then this procedure is repeated until we have updated $N$ nodes in a single time step.

Recently scale-free distributions came up in the arena of $RBNs$
 in order to match biological network scenarios. The first works  have only  used 
computer simulations \cite{castro2004scale,iguchi2007boolean,novikov2008regulatory}. Later, some authors have focused on analytical approaches \cite{drossel2009critical,aldana2003boolean,lee2008broad,zhou2005dynamic}
giving a more detailed insight of the problem. 
In this work we study  random Boolean dynamics in  scale-free networks.
In our model  we have nodes with Boolean variables $(0,1)$. The connections among 
nodes obey the topological structure of scale-free networks and the relation 
among nodes variables is performed via randomly chosen Boolean functions. We 
suppose that the dynamics is driven only by XOR and AND functions, which appear for each node with
 probabilities $p$ and $1-p$, respectively. 
We chose the AND and XOR logical functions in order to simplify the
model, avoiding the necessity of defining $2^{2^K}$ different Boolean relations for
each node. Note that any Boolean function can be written as a linear combination of AND, XOR and OR. The chosen functions are examples of extreme cases. When the AND function is applied to a set with $K$ Boolean variables,
we will obtain 1 only if {\sl all} variables are 1's. On the other hand, when the XOR function is applied to the same set,
we will obtain 1 if the number of 1's of the set is odd. These observations imply that AND represents a very selective
dynamical rule, only one configuration of the $2^K$ possible ones furnishes 1 as output, while XOR is related to a non-selective
dynamical one because half of the set configurations give 1 as output.
Moreover, it is
known from previous works that the AND function leads to an ordered regime with two
fixed points, where all variables are 0's or 1's, and that the dynamical behavior generated
by the XOR function is more complex.
Since this kind of functions are the
Boolean counterparts of real reactions in cell regulatory system \cite{kau1,wei},
 they  are important to the study of biologic networks.
We choose the parallel mode as updated method.
In order to compare with another models, and for technical reasons, we also study such dynamics in networks without a scale-free topology. The Kauffman NK model is briefly discussed as well.
We performed two distinct types of dynamics. In the 
first case the  node that will be updated is regulated only by the nodes which are connected to it. Rarely the node is connected to itself. It means that its state is 
defined  by the state of its neighbors, and its own state is almost never taken into account. 
In the second case we explore self-regulation, which means that the new state of the updating node is defined by its neighbors and 
always for its own state. We choose to study such case because self-regulation is a well know feature of 
genetic regulatory networks   \cite{guelzim2002topological}.
In section II we introduce the scale-free networks and the Boolean dynamics used in
this work. The numerical simulations of the scale-free networks are discussed in  section III.
In section IV we present a mean-field (annealed)
approximation for these dynamics, leading to an analytical way to calculate 
the average density of $1's$ and the Hamming distance. We apply the annealed
approximation to networks without a scale-free topology, to the Kauffman NK model and to scale-free networks.
The comparison between the numerical results and those of the annealed approximation is presented in section V.
We summarize our results in the last section. 

\section{Networks and the dynamics} 

 We have generated two classes of distinct networks, classified by the smallest number 
of links that a node can have, in other words, by the smallest possible connectivity $k_{min}$ of the network. These networks were grown by the {\it Growing Network with Re-direction} algorithm \cite{krapivsky2001organization} and can
be classified as

\begin{itemize}
\item $k_{min}=1$ -- a node of the network, called old node, is selected with 
uniform probability;  then a new node  is linked to it  
with probability $1-r$ or it is redirected
to the ancestor of the old node with probability $r$;

\item $k_{min}=2$ -- a new node has two links; the first link with 
 an old node, selected with uniform probability,  
 is established with probability $1-r$ or it is redirected
to one of the two ancestors of the old node with probability $r$; 
 the same procedure is repeated  for the second link.
\end{itemize}
For $k_{min}=1$, initially we have three nodes cyclically connected (the ancestors of nodes 1, 2 and 3 are 3, 1 and 2 respectively).
A new node is randomly connected  to an old node (one of the three initial nodes). Then, this new link can be redirected to the ancestor of the old node with probability $r$. This growing algorithm is repeated until we have $N$ nodes in the network.
When $k_{min}=2$, each one of the initial three nodes has the other two nodes as ancestors. Now, a new node is randomly connected to two
 old nodes, and each new link can be redirected to the ancestor with probability $r$. We repeat this procedure until we have a
network with $N$ nodes. These algorithms create 
scale-free networks characterized by a connectivity distribution $P(k)\sim k^{-\gamma}$
with $\gamma=r^{-1}+1$ \cite{krapivsky2001organization}. When $r=0$, the nodes are linked in
a entirely random fashion and for $r=1$, all nodes are connected to one of the three
initial nodes (super-hubs). The linear preferential attachment model of Barab\`asi and Albert is obtained for $r=0.5$ ($\gamma=3$).
In this work, we consider three typical values of $r$: $r=0.5$ (the Barab\`asi and Albert model),  $r=0.8$, representing
models with large hubs and $r=0.35$, representing models with small hubs.

A logical variable $\sigma_i(t)$ is assigned to each node $i$ and the 
state of the network at time $t$ is represented by a set of Boolean variables
$(\sigma_1(t), \sigma_2(t), \sigma_3(t), ..., \sigma_N(t))$. Each variable
$\sigma_i(t)$ is controlled by $k_i$ elements of the network 
 $\{\sigma(t)\}_{k_i}=\{\sigma_{i_1}(t), \sigma_{i_2}(t), ..., \sigma_{i_{k_i}}(t)\}$. 
 If $k_{min}=2$, $k_i$ is the connectivity of $i$-th node
and the control elements are the nodes connected to it. When $k_{min}=1$, we have nodes with
only one link. Since we need two inputs to apply the Boolean functions, it is natural to
assume that the node itself must always participate of the dynamics.
Now, the control
elements are the $i$-th node itself and nodes connected to it. It means that
each node has an extra link to itself (self-regulation). In this case, $k_i$ is
the connectivity plus 1. The dynamics is given by
\begin{equation}
\sigma_i(t+1)=G_i(\{\sigma(t)\}_{k_i}),
\label{eq1}
\end{equation}
where $G_i$ is the random function
\begin{equation}
G_i = {\begin{cases} \textrm{AND}\,~\textrm{with probability}\ $1-p$ \\
              \textrm{XOR} \,~\textrm{with probability}\ $p$,\end{cases}}
\label{eq2}
\end{equation}
that is assigned to each node $i$. Here $p$ is an external parameter that controls how the logical functions  AND and
 XOR are distributed in the network.

An initial state $\{\sigma(0)\}$ is created by assigning randomly 0's and 1's to 
all nodes.
 A damaged copy $\overline{\{\sigma(0)\}}$ of the initial state is also created  
  by changing the value of only one randomly chosen node. Since the Hamming distance
of two configurations is the number of nodes that have different values in these configurations,
the  Hamming distance between $\{\sigma(0)\}$  and $\overline{\{\sigma(0)\}}$ is 1. 
  Both the initial state and  its copy evolve under the control of equation (\ref{eq1}).
 Once the new state of all nodes is calculated the entire network is
updated (synchronous update) and the system goes to the next Monte Carlo time step (mcs).

We characterize the dynamical behavior by the
average density of 1's
$$
M(p,t)=\lim_{N\to\infty} \left\langle\frac{1}{N}\sum_{i=1}^N\sigma_i(t)\right\rangle,
$$
and by the average of the Hamming distance
$$
D(p,t)=\lim_{N\to\infty} \left\langle\frac{1}{N}\sum_{i=1}^N|\sigma_i(t)-\overline{\sigma_i(t)}|
\right\rangle.
$$
 Here, $\langle\ldots\rangle$ is an average over the
 initial states ($\{\sigma(0)\}$, $\overline{\{\sigma(0)\}}$), and over  sets of links of 
 a grown network with a specific $\gamma$ and with the same $p$.
 
 After a  transient time these quantities reach the stationary values $M(p)$ and $D(p)$ that 
 can be defined as 
\begin{equation}
M(p)= \lim_{T \to \infty} \frac{1}{T} \int_t^{t+T} M(p,t^\prime) dt^\prime,
\label{eq5}
\end{equation}
\begin{equation}
D(p)= \lim_{T \to \infty} \frac{1}{T} \int_t^{t+T} D(p,t^\prime) dt^\prime.
\label{eq6}
\end{equation}

\section{Numerical Simulations}

\subsection{Data and results for $k_{min}=1$}

In order to consider finite size effects, we grew networks with $N=1\times 10^4$,
$N=2\times 10^4$ and  $N=4\times 10^4$ nodes.  The  averages were performed with a number of
samples varying from  $10^2$ (large $p$ and large $N$) up to
 $5\times 10^4$ (small $p$ and small $N$). The probability $p$ was taken in the
 interval $[0.001,\, 0.8]$ for the three values of $r$ ($0.35$, $0.5$ and $0.8$).
  Since for $p<0.001$ the averages quantities were very small, we decided that a good lower limit was $p=0.001$.
 The upper limit $p=0.8$ was chosen because the values of the average quantities were similar,
in a log-log scale.
The stationary values, $M(p)$ and $D(p)$, were reached after a very short transient time 
 ($20$ mcs). Estimations of $M(p)$ and $D(p)$, which are defined in Eqs. (\ref{eq5}) and (\ref{eq6}), were performed by considering $t=20~ mcs$ and $T=80~ mcs$. These stationary values remain basically the same if we increase both $t$ and $T$.

\begin{figure}
\centering
{\includegraphics[width=8cm,height=7cm]{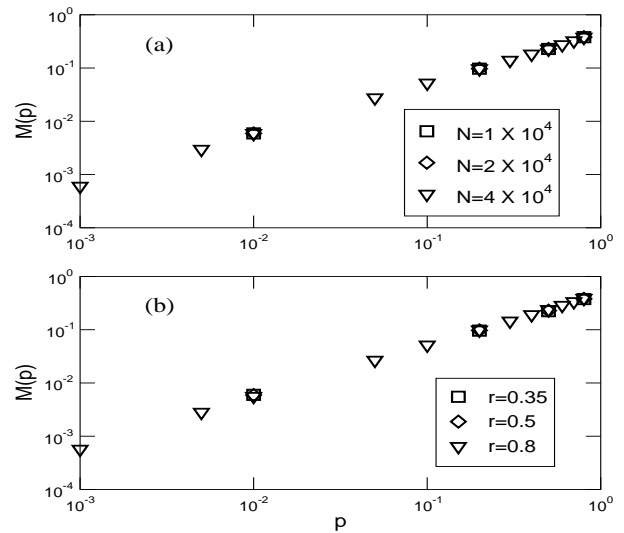}}
\caption{Log-log plots of $M(p)~\textrm{vs}~p$  for: (a) a network with $k_{min}=1$, $r=0.5$ and different $N$
 (b) a network with $k_{min}=1$, $N=10^4$ and different $r$}
\label{fig2}
\end{figure}

\begin{figure}
  \centering
 {\includegraphics[width=8cm,height=6cm]{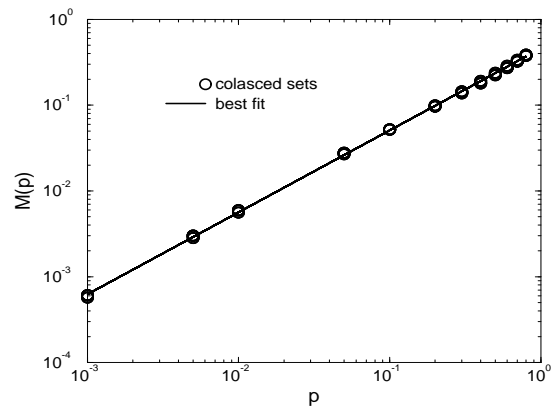}}
\caption{Log-log plot of $M(p)~\textrm{vs}~p$  for $k_{min}=1$, $N$ varying from 
$1\times 10^4$ up to $4\times 10^4$  and $r=0.35,\, 0.5$ and $0.8$ . It shows 
the best fit of the coalesced sets.}
\label{fig3}
\end{figure}
 
We can see from Fig.~\ref{fig2} that $M(p)$ behaves as a power function of type $p^m$ for the entire range of $p$.
 By comparing the behavior of
the smallest network ($N=1\times 10^4$) with the largest one ( $N=4\times 10^4$)  we can
see that finite size effects are small for $M(p)$. Moreover, it seems that the exponent $m$
does not depend on $r$.
 In order to evaluate the
exponent $m$ we coalesce all different sets ($N$ varying from $10^4$ up to $4\times 10^4$ and 
$r$ from 0.35 up to 0.8) and we do a best fit.
 This is shown in Fig.~\ref{fig3}. We obtain that
$$
M(p)= ap^m,
$$
with $a=0.46\pm 0.01$ and $m=0.96\pm 0.01$.
 Note that we have evaluated the exponent by considering
 approximately 2 orders of magnitude in the $p$ variable and that the fit is very good. In fact,
 in all fitted data, we obtained a correlation coefficient larger that $0.999$. 
  
\begin{figure}
\centering
{\includegraphics[width=8cm,height=7cm]{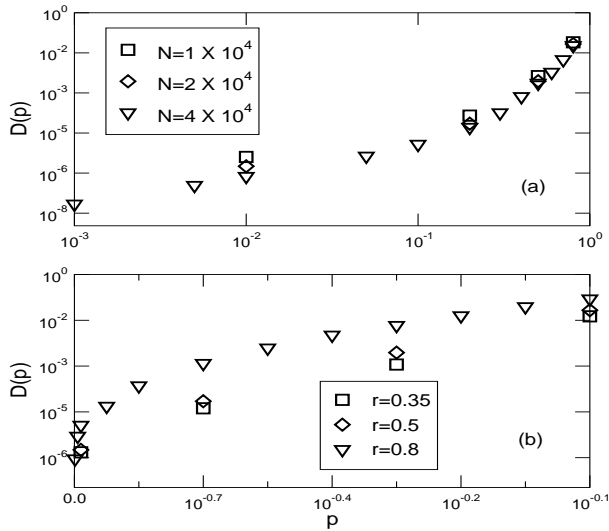}}
\caption{(a) Log-log plot of $D(p)~\textrm{vs}~p$  for a network with $k_{min}=1$, $r=0.5$ and different $N$;
         (b) Linear-log plot of $D(p)~\textrm{vs}~p$ for a network with $k_{min}=1$, $N=2\times 10^4$ and different $r$.}
\label{fig4}
\end{figure}

Plots of $D(p)$ versus $p$ are shown in Fig.~\ref{fig4} for $r=0.5$ and networks with
different sizes.
 We can see  that $D(p)$ has a power law behavior ($D(p)\sim p^d$) only when the probability $p$ is close to $p=0$. 
 Outside of these region, $D(p)$ grows exponentially. By comparing the behavior of networks 
 with different sizes, we observe that finite size effects are now important. 

\begin{figure}
\centering
{\includegraphics[width=8cm,height=7cm]{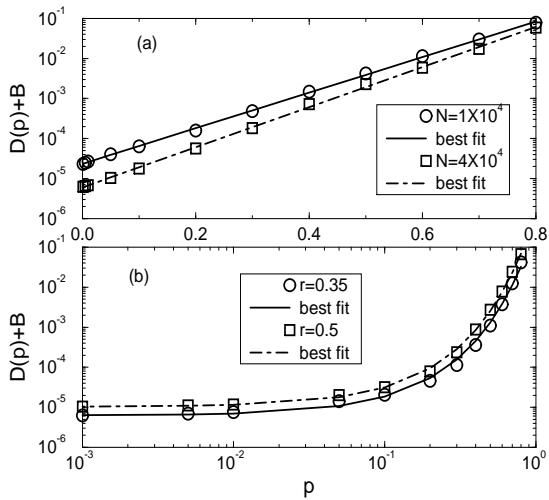}}
\caption{(a) Linear-log plot of $D(p)+B~\textrm{vs}~p$  for a network with $k_{min}=1$, $r=0.5$ and different $N$;
         (b) Log-log plot of $D(p)+B~\textrm{vs}~p$ for a network with $k_{min}=1$, $N=2\times 10^4$ and different $r$.}
\label{fig5}
\end{figure}
It turns out that all results can be well fitted by 
$$
D(p)= B \exp (C p) - B,
$$ 
with $C$ and $B$ depending on the size of the network and on the parameter $r$. This can
be seen in Fig.~\ref{fig5}. When $p\approx 0$, we have that $D\sim BCp$. 

\subsection{Data and results for $k_{min}=2$} 

The simulation for the networks with $k_{min}=2$ were realized in the same 
way that for $k_{min}=1$, and the probability $p$ was taken in the interval 
$[0.01, 0.9]$ for $r=0.2$, $0.5$ and $0.8$. The range of $p$ is rather narrow 
in this case, since the dynamics for $k_{min}=2$ is  more sensible to $p$ values, 
leading to $M(p,t) \rightarrow 0$ and $D(p,t) \rightarrow 0$ when $p \sim
0.001$. This behavior perhaps is  related to the one found  for RBN with only part of the canalyzing functions 
as update functions, $K=2$, and
small $p_+$, where $p_+$ is the probability that a connection be excitatory \cite{greil-drossel}.
\begin{figure}
\centering
{\includegraphics[width=8cm,height=6cm]{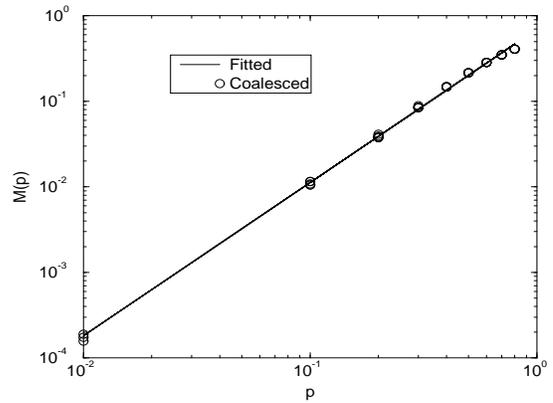}}
\caption{Log-log plot of $M(p)~\textrm{vs}~p$  for $k_{min}=2$, $N$ varying from 
$1\times 10^4$ up to $4\times 10^4$  and $r=0.2,\, 0.5$ and $0.8$ . It shows 
the best fit of the coalesced sets.}
\label{fig7}
\end{figure} 
Fig. \ref{fig7} is similar to fig. \ref{fig3}, where all sets of $r$ and $N$ are 
coalesced in one plot. We can see that the finite size effect is 
very small and we have 
$$
M(p) = ap^m,
$$
with $a=0.70\pm 0.02$ and $m=1.79\pm 0.01$.

\begin{figure}
\centering
{\includegraphics[width=8cm,height=6cm]{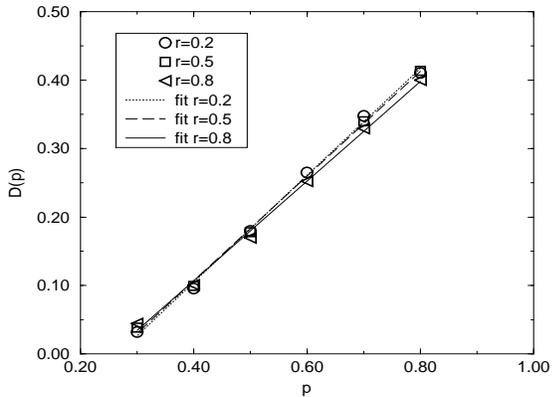}}
\caption{Linear plot of $D(p)~\textrm{vs}~p$  for $k_{min}=2$,
$r=0.2$, $0.5$ and $0.8$ for $N=10^3$.}
\label{fig8}
\end{figure}

As we can see in Fig. \ref{fig8}, the Hamming distance $D(p)$ follows 
a linear dependence with $p$ for large values of $p$
(the plot shows $p \ge 0.3$). In this region the behavior of $D(p)$ is 
almost independent of $r$, and we have:
$$
D(p) \sim ap,
$$
where $a=0.78\pm 0.02$, $0.76\pm 0.02$ and $0.72\pm 0.02$ for $r=0.2$, $0.5$ and $0.8$ 
respectively. However, the behavior of $D(p)$ for $p \le 0.3$ does not 
provide any suitable fit since for low values of $p$ we have a high 
concentration of AND function and a low concentration of XOR function, 
leading most of nodes to the $0$ state. 

Note that results discussed in this section are  valid for other values of $r$. We discuss
only the cases $r=0.5$ (linear preferential attachment), one case with $r<0.5$ ($r=0.35$ or $r=0.2$)
and one case with $r>05$ ($r=0.8$ or $r=0.7$) because they represent typical behaviors.
 We have simulated other cases, with less samples, and the results are similar.
 Although we present $D(p,t)$ for the initial condition $D(p,0)=1$, we have also simulated
cases with $D(p,0)>1$. We find similar results, probably because this new initial condition 
is a later state of the initial condition with the smallest Hamming distance.
In the next section we will develop a mean-field approach in order to see if its results
 agree with the numerical ones just obtained.

\section{Mean-field  approximation}

In this section we present a mean-field approach (MF). It is based on a work of Derrida and Pomeau \cite{derr},
in which an annealed approximation was done for the Kauffman $NK$ model. The NK model is a cellular 
automaton with $N$ nodes holding logical variables. Each node $\sigma_i$ is connected with 
another $K$ nodes of the network meaning that all nodes have the same connectivity $K$.
The dynamics is given by Eq. (\ref{eq1}) where $G_i$ is a random Boolean function. Although
the model is defined by  quenched disorder, i.e, the Boolean function $G_i$  and the $K$ nodes connected
to each node $\sigma_i$ are only randomly chosen at the initial time, in the Derrida and Pomeau approximation it is assumed an annealed disorder. It means that the Boolean function  and the $K$ nodes are randomly chosen at each time step. Moreover, in such approximation
the effect of the Boolean functions on a node is described by  probabilities that the output be 0 and 1. The approximation to our
problems is similar to that of Derrida and Pomeau. However, a difference appears in the application of the Boolean functions. Instead of using a probabilistic description for the effect of the Boolean functions,
 we  determine the effect of applying the XOR and AND operators in each configuration. We will first discuss the case $k_{min}=2$
because it is more illustrative than the more simple case $k_{min}=1$.

\subsection{Average density of $1$'s for $k_{min}=2$}\label{subsec-jaff1}

Let us start our evaluation for the model without a scale-free topology. The dynamics is given by  equations (\ref{eq1}) and (\ref{eq2}). 
Suppose that the  configuration of the system at time $t$, $\{\sigma_i(t)\}$, consists of $n$ nodes with $\sigma=1$ 
and $N-n$ nodes with $\sigma=0$.
In order to study the density 
of $1$'s, we can separate the configuration  $\{\sigma_i(t)\}$  in two sets: (i) set
$\mathcal{A}(t)$ where all the nodes have $\sigma=0$, and (ii) set $\mathcal{B}(t)$ 
where all the nodes have $\sigma=1$. 
 Now we must randomly choose $K$ nodes that are linked
to the node $i$.
 The probability 
 that a given link comes from set $\mathcal {A}(t)$ is $1-x=(N-n)/N$ and the probability that it comes  
from $\mathcal {B}(t)$ is $x=n/N$. Since each node has $K$ links, the list $P_K$ of the probabilities of the  possible link configurations is
\begin{eqnarray}
P_K=\left\{ {K \choose K}(1-x)^K,~~ {K \choose K-1}(1-x)^{K-1}x,~ \right. \label{jafflist1} \\
\left. {K \choose K-2}(1-x)^{K-2}x^2 ,~~ ... , {K \choose 0}x^K \right\} . \nonumber
\end{eqnarray}
 We must now evaluate the output of each
possible configuration under the application of operators AND and XOR.
The Boolean function AND generates an output $1$ if all inputs come from set $\mathcal {B}(t)$. 
The probability of this configuration is given by the last term of the list (\ref{jafflist1}). The function 
XOR, however, produces $1$ as output when its number of inputs equal to $1$ is an odd 
number, meaning that the number of links coming from set $\mathcal {B}(t)$ is odd. 
Therefore the probability $X$ of obtaining $1$ as output is
$$
X=p\sum_{ \substack{m=1 \\m~odd} }^{K}{K\choose m}(1-x)^{K-m}x^m +qx^K,
$$
where $p$ and $q=1-p$ are the probabilities related to the XOR and AND operators. Since we have that 
$$
(y\pm x)^K=\sum_{m=0}^K {K\choose m}y^{K-m}(\pm x)^m,
$$
the sums of the even and odd terms are given by
\begin{eqnarray}
\sum_{\substack{m=0\\m~even}}^K {K\choose m}y^{K-m} x^m&=&\frac{(y+x)^K+(y-x)^K}{2},\label{eq10}\\
\sum_{\substack{m=1 \\m~odd}}^K {K\choose m}y^{K-m} x^m&=&\frac{(y+x)^K-(y-x)^K}{2}.\label{eq11}
\end{eqnarray}
 It follows that
$X=\frac{p}{2}[1-(1-2x)^K]+qx^K$.
Assuming homogeneity, we can identify
$X$ as $M(p,t+1)$, the fraction of $1$'s at time $t+1$.
Then we obtain that $M(p,t+1)$ depends only on $M(p,t)$ as
\begin{equation}
M(p,t+1)=\frac{p}{2}\{ 1-[1-2M(p,t)]^K\} +q{M(p,t)}^K.\label{eq13}
\end{equation}

Finally let us consider the scale-free topology. Each node now has $k$ links with
probability $P(k)$. Therefore the above equation can be written as
\begin{eqnarray}
M(p,t+1)&=&\frac{p}{2}\left\{ 1-\sum_{k=2}^\infty P(k)[1-2M(p,t)]^k\right\} \nonumber \\
&+&q\sum_{k=2}^\infty P(k){M(p,t)}^k.\label{eq15}
\end{eqnarray}

\subsection{Average Hamming distance for $k_{min}=2$}\label{subsec-jaff2}

The calculation of the Hamming distance is done in a similar way. 
Let us again first study the model without scale-free topology. At time $t$ we are interested in the configuration $\{\sigma (t)\}$ resulting from the evolution of the initial configuration,  and in the configuration  $\{\overline{\sigma(t)}\}$, which appears from the evolution of the initial damaged copy. 
Suppose that they differ by $n$ nodes.
Following Derrida and Pomeau   \cite{derr}, we define  
 two sets: $\mathcal{E}(t)$ and $\mathcal{F}(t)$. The first one is the set of all
nodes of $\{\sigma(t)\}$ and $\{\overline{\sigma(t)}\}$ that have the same values.  The set $\mathcal{F}(t)$
is composed by the $n$ nodes that have different values in the two configurations.
 Therefore the nodes which have all links coming from set $\mathcal{E}(t)$ 
will have the same values at time $t+1$ in the $\{\sigma(t)\}$ and $\{\overline{\sigma(t)}\}$ configurations and they will not contribute to the Hamming distance.
 On the other hand, the Hamming distance could be 
changed by the nodes which have at least one link coming from the $n$ nodes of $\mathcal{F}(t)$.
 
Let us also define $E_0$ and $E_1$ as the number of nodes in the set $\mathcal{E}(t)$ with
the values $0$ and $1$, respectively. $F_0$ is the number of nodes in the set $\mathcal{F}(t)$ that have
$\sigma (t)=0$, and $F_1$ is the number of nodes with $\sigma (t)=1$.
Observe that $F_0+F_1=n$ and $E_0+E_1=N-n$.
The next step is to focus in a particular node $i$ and to determine the probability of have $K$ randomly chosen nodes  linked to it.
 The probability that a link comes from set $\mathcal{E}(t)$ with the corresponding node
having value $1$ is  $z_1=E_1/N$. If the link comes from the same set but the node of $\mathcal{E}(t)$
has value $0$, the probability will be $z_0=E_0/N$. $w_0=F_0/N$ and $w_1=F_1/N$ are the probabilities that a link comes from $\mathcal{F}(t)$ when the corresponding elements of the set have values $0$ and $1$, respectively.
 Since $E_0+E_1+F_0+F_1=N$, it is obvious that $z_0+z_1+w_0+w_1=1$.

We are interested in the evaluation of $W_1$, the probability that $\sigma_i(t+1)=1$ and 
$\overline{\sigma+_i(t+1)}=0$, and of $W_0$, the probability that $\sigma_i(t+1)=0$ and 
$\overline{\sigma_i(t+1)}=1$. 
The first step is to study the situation in which the node $i$ has $K-1$ links in set $\mathcal{E}(t)$ and
only one link in $\mathcal{F}(t)$. The list of the probabilities of the possible link configurations is
\begin{eqnarray}
P_1^{(1)}&=&\left\{ {K \choose 1}w_0,~{K \choose 1}w_1\right\} \left\{ z_0^{K-1},~{K-1 \choose 1} z_0^{K-2}z_1,\right.
\nonumber\\
&&\left. {K-1 \choose 2} z_0^{K-3}z_1^2,~\ldots,~z_1^{K-1}\right\},\label{jafflist2}
\end{eqnarray}
where we must multiply each element of the first list by each element of the second one.
 We must now evaluate the output of each
possible configuration under the operator XOR. 
For the first configuration (${K \choose 1}w_1 z_0^{K-1}$), the XOR operation furnishes that
$\sigma_i=1$ in configuration $\{\sigma(t+1)\}$ and $\sigma_i=0$ in $\{\overline{\sigma(t+1)}\}$,
 implying that $p{K \choose 1}w_1 z_0^{K-1}$ will contribute to $W_1$. Note that the extra probability $p$ is related to the XOR operator.
The second configuration (${K \choose 1}w_1{K-1 \choose 1} z_0^{K-2}z_1$) will contribute to
$W_0$ because the application of XOR give-us that $\sigma_i=0$ in  $\{\sigma(t+1)\}$ and 
$\sigma_i=1$ in $\{\overline{\sigma(t+1)}\}$. Since the third term ($ {K \choose 1}w_1{K-1 \choose 2} z_0^{K-3}z_1^2$) will contribute to $W_1$, it is easy to infer that for configurations beginning with $w_1$, the
terms $z_1^m$ with $m$ even contribute to $W_1$ and the ones with $m$ odd enter in $W_0$. A similar
analysis shows that for configurations beginning with $w_0$, the
terms  with $m$ even contribute to $W_0$ and the ones with $m$ odd enter in $W_1$.
When we apply the AND operator, only the terms $w_1z_1^{K-1}$ and $w_0z_1^{K-1}$ give
 no null contributions to $W_1$ and $W_0$, respectively. Therefore the contributions of
the list (\ref{jafflist2}) to $W_1$ and $W_0$ can be written as
\begin{eqnarray*}
W_1^{(1)}&=&{K \choose 1}\left[p w_1\sum_{\substack{m=0\\m~even}}^{K-1}{K-1 \choose m}z_0^{K-m}z_1^m\right.\\
                            &&\left. +pw_0\sum_{\substack{m=1\\m~odd}}^{K-1}{K-1 \choose m}z_0^{K-m}z_1^m
           +qw_1z_1^{K-1}\right],
\end{eqnarray*}
\begin{eqnarray*}
W_0^{(1)}&=&{K \choose 1}\left[ pw_0\sum_{\substack{m=0\\ m~even}}^{K-1}{K-1 \choose m}z_0^{K-m}z_1^m\right.\\
                        &&\left. +pw_1\sum_{\substack{m=1\\ m~odd}}^{K-1}{K-1 \choose m}z_0^{K-m}z_1^m
           +qw_0z_1^{K-1}\right].
\end{eqnarray*}
Using Equations (\ref{eq10}) and (\ref{eq11}), $W_1^{(1)}$ can be written as
\begin{eqnarray*}
W_1^{(1)}&=&{K \choose 1}\left[\frac{p}{2}(w_1+w_0)(z_0+z_1)^{K-1}\right. \\
&&\left. +\frac{p}{2}(w_1-w_0)(z_0-z_1)^{K-1}
                             +qw_1z_1^{K-1}\right].
\end{eqnarray*}
The equation for $W_0^{(1)}$ is obtained by changing $w_0$ by $w_1$, and $w_1$ by $w_0$
in the above equation.

The second step is to study the situation in which the node $i$ has $K-2$ links in set $\mathcal{E}(t)$ and
two links in $\mathcal{F}(t)$. Now, the list of the probabilities of the possible link configurations is
\begin{eqnarray}
&&P_1^{(2)}={K \choose 2}\left\{w_0^2,~2w_0w_1,~w_1^2\right\}\left\{ z_0^{K-2},
\right.\\&&\left.{K-2 \choose 1} z_0^{K-3}z_1,
~{K-2 \choose 2} z_0^{K-4}z_1^2,~\ldots,~z_1^{K-2}\right\}.\nonumber\label{jafflist3}
\end{eqnarray}
There is no contribution of the XOR operator. The AND operator furnishes that only 
two configurations ($w_1^2z_1^{K-2}$ and $w_0^2z_1^{K-2}$), multiplied by probability $q$,
contribute to $W_1$ and $W_0$. Therefore we find that
$$
W_1^{(2)}={K \choose 2}qw_1^2z_1^{K-2},~~{\rm and}~~W_0^{(2)}={K \choose 2}qw_0^2z_1^{K-2}.
$$

The third step is to evaluate the probabilities generated by the application of XOR and
AND in the situation in which the node $i$ has $K-3$ links in set $\mathcal{E}(t)$ and
three links in $\mathcal{F}(t)$. This situation is similar to the first one. We obtain
that
\begin{eqnarray*}
W_1^{(3)}&=&{K \choose 3}\left[\frac{p}{2}(w_1+w_0)^3(z_0+z_1)^{K-3}\right.\\
&&\left. +\frac{p}{2}(w_1-w_0)^3(z_0-z_1)^{K-3}
                             +qw_1^3z_1^{K-3}\right].
\end{eqnarray*}
The result for $W_0^{(3)}$ is identical with the previous one if we change 
$w_0$ by $w_1$, and $w_1$ by $w_0$.

The fourth step is similar to the second one, and so on. Since $W_1=\sum_{m=1}^K
W_1^{(m)}$, we have that
\begin{eqnarray*}
&&W_1=\frac{p}{2}\sum_{m,odd}^K{K \choose m}\left[(w_1+w_0)^m(z_0+z_1)^{K-m}\right.\\
 &&\left. +(w_1-w_0)^m(z_0-z_1)^{K-m}\right]
                             +q\sum_{m=1}^K{K \choose m}w_1^mz_1^{K-m}.
\end{eqnarray*}
To obtain $W_0$ we substitute $w_0$ by $w_1$, and $w_1$ by $w_0$ in this equation.
Using again Eq. (\ref{eq10}), we obtain that
\begin{eqnarray}
W_1&=& q[(w_1+z_1)^K-z_1^K] +\frac{p}{4}\{1-[1-2(w_1+w_0)]^K\nonumber\\ 
&+&[1-2(z_1+w_0)]^K-[1-2(z_1+w_1)]^K\},\label{eq21}
\end{eqnarray}
 
\begin{eqnarray}
W_0&=& q[(w_0+z_1)^K-z_1^K] +\frac{p}{4}\{1-[1-2(w_1+w_0)]^K\nonumber\\ 
&+&[1-2(z_1+w_1)]^K-[1-2(z_1+w_0)]^K\}.\label{eq22}
\end{eqnarray}
Assuming homogeneity, we can identify
$W_1$ as $w_{1,t+1}$, the fraction of $1$'s of set $\mathcal{F}(t)$ at time $t+1$,
and $W_0$ as $w_{0,t+1}$. Note that the fraction of $1$'s of the system is
given by $M(p,t+1)=z_{1,t+1}+w_{1,t+1}$ and that it was already evaluated (see Eq.(\ref{eq13})).
Identifying $z_{1,t+1}$ with $Z_1$, we obtain that
$$
Z_1=M(p,t+1)-W_1.
$$
Observe that the equations for $W_1$, $W_0$ and $Z_1$ describe completely our system,
since the equation for $Z_0$ is obtained from the normalization condition.  However
it is usual to work with variables $M(p,t+1)$, the density of $1$'s and with $D(p,t+1)$, the Hamming distance.
From the definition of the Hamming distance we have that $D(p,t+1)=w_{1,t+1}+w_{0,t+1}$,
implying that
\begin{eqnarray*}
D(p,t+1)&=&q\{ M(p,t)^K+[D(p,t)-M(p,t)+2z_{1,t}]^K\\
&-&2z_{1,t}^K\} +\frac{p}{2}\{ 1-[1-2D(p,t)]^K\}.
\end{eqnarray*}
The equations for $D(p,t+1)$, $M(p,t+1)$ and $z_{1,t+1}$ also describe completely
the system. However, if $w_1=w_0$ we can see from Eqs. (\ref{eq21}) and (\ref{eq22})
that $W_1=W_0$. Solving numerically these equations, we obtain that each initial configuration with $w_1\not= w_0$ evolves to
a fixed point with $W_1=W_0$. 
Then we can assume that $w_0=w_1$, without loss of generality, and the dynamics of the system is described by only two equations, namely
\begin{eqnarray}
D(p,t+1)&=&2q\{ M(p,t)^K-[M(p,t)-\frac{D(p,t)}{2}]^K\}\nonumber
 \\
&+&\frac{p}{2}\{ 1-(1-2D(p,t))^K\} ,\label{eq25}\\
M(p,t+1)&=&qM(p,t)^K \label{eq26}\\
&+&\frac{p}{2}\{ 1-[1-2D(p,t)]^K\}.\nonumber
\end{eqnarray}

Finally let us consider the model with the scale-free topology. Since each node now have $k$ links with
probability $P(k)$, the above equations, which are valid for $w_1=w_0$, can be written as

\begin{eqnarray}
D(p,t+1)&=&2q\left\{ \sum_{k=2}^\infty P(k)\{ M(p,t)^k\right.\nonumber\\
&-&\left.[M(p,t)-\frac{D(p,t)}{2}]^k\}\right\}\nonumber\\
&+&\frac{p}{2}-\frac{p}{2}\sum_{k=2}^\infty P(k)[1-2D(p,t)]^k,\label{eq27}
\end{eqnarray}
\begin{eqnarray}
M(p,t+1)&=&\frac{p}{2}\left\{ 1-\sum_{k=2}^\infty P(k)[1-2M(p,t)]^k\right\} \nonumber\\
&+&q\sum_{k=2}^\infty P(k){M(p,t)}^k.\label{eq28}
\end{eqnarray}

\subsection{$M(p,t)$ and $D(p,t)$ for $k_{min}=1$}

The main difference between this case and the previous one is the self-regulation mechanism: the node itself
always participates in its own dynamics. Itself and the $K$ nodes connected to it are
the control elements of the dynamics. Following a similar procedure of the subsection \ref{subsec-jaff1},
we obtain that
\begin{eqnarray}
M(p,t+1)&=&\frac{p}{2}\left\{ 1-\sum_{k=1}^\infty P(k)[1-2M(p,t)]^{k+1}\right\}\nonumber\\
 &+&q\sum_{k=1}^\infty P(k){M(p,t)}^{k+1},
\label{eq30}
\end{eqnarray}
where $P(k)$ is the probability that a node had $k$ links. 

The evaluation of the Hamming distance follows similar steps of subsection \ref{subsec-jaff2}.
 For scale-free systems we again have the quantities $W_0$, $W_1$ and $Z_1$. It turns out that when
$w_1=w_0$ we have $W_1=W_0$. When $w_1=w_0$, the dynamics of the system is given by Eq. (\ref{eq30})
and by
\begin{eqnarray}
D(p,t+1)&=&2q\left\{ \sum_{k=1}^\infty P(k)\{ M(p,t)^{k+1}\right.\nonumber\\
 &-&\left.[M(p,t) -\frac{D(p,t)}{2}]^{k+1}\}\right\}\nonumber\\
&+&\frac{p}{2}-\frac{p}{2}\sum_{k=1}^\infty P(k)[1-2D(p,t)]^{k+1}.
\label{eqnew1}
\end{eqnarray} 
If we put $P(k)=\delta_{k,K}$  in Eqs. (\ref{eq30}) and (\ref{eqnew1}), we obtain the results for the model without  a scale-free topology. They are  similar to the ones obtained
for the case without self-regulation, but with $K$ replaced by $K+1$ (see Eqs. (\ref{eq25}) and (\ref{eq26})).
For scale-free systems, the dynamics with self-regulation is similar to the usual dynamics if we change $k$ by $k+1$, except
in the  distribution of connectivity $P(k)$ (see Eqs. (\ref{eq27} and \ref{eq28}). 
Therefore the dynamics with self-regulation is different from the usual case. Let us investigate if this fact is
also true for the  Kauffman NK model.

\subsection{Kauffman model with self-regulation}

The Kauffman NK model consists of $N$ nodes holding logical variables $\sigma_i$. Each node is connected with   any $K$ nodes of the network. Observe that a node $i$ can have  a link to itself with 
small probability ($K/N$).
The dynamics, given by Eq. (\ref{eq1}), is determined by a random Boolean function $G_i$. 
In the Derrida and Pomeau annealed approximation   \cite{derr}, the Boolean function  and the $K$ nodes are randomly chosen at each time step. The configuration $\{\sigma(t)\}$ is split in the sets $\mathcal{F}(t)$, which consists of  nodes having different values of 
$\sigma$ in configurations
$\{\sigma (t)\}$ and $\{\overline{\sigma(t)}\}$, and $\mathcal{E}(t)$ when the previous condition does not hold. Then we are able to define the probabilities $w$ and $z$ that a link of a particular node comes
from sets $\mathcal{F}(t)$ and $\mathcal{E}(t)$, respectively. Obviously we have that $z+w=1$. We want to evaluate the probability $W$ that  node $i$ will have different values in $\{\sigma (t+1)\}$ and 
$\{\overline{\sigma(t+1)}\}$. If all $K$ links came from $\mathcal{E}(t)$, $\sigma_i(t+1)$ will have the same value in both $\{\sigma (t+1)\}$ and $\{\overline{\sigma(t+1)}\}$.
However, if at least one link comes from $\mathcal{F}(t)$, $\sigma_i(t+1)$ has a positive probability of having
different values in $\{\sigma (t+1)\}$ and $\{\overline{\sigma(t+1)}\}$. 
Due to the random Boolean function assignment any node can be $0$ or $1$ with probability $1/2$, and the probability that 
$\sigma_i(t+1)$ will have different values in $\{\sigma (t+1)\}$ and $\{\overline{\sigma(t+1)}\}$ is $1/2$.
Therefore the probability $W$ is given by
$$
W=\frac{1}{2}\left[{K \choose 1}z^{K-1}w+{K \choose 2}z^{K-2}w^2+\ldots + w^K\right].
$$
Assuming that the system is homogeneous, we can identify the fraction of nodes with different
values in $\{\sigma (t+1)\}$ and $\{\overline{\sigma(t+1)}\}$, $W$, with the Hamming distance
$D(p,t+1)$. Using that $w=D(p,t)$ in the previous equations we obtain the traditional equation of Derrida and Pomeau   \cite{derr}, namely 
$$
D(p,t+1)=\frac{1}{2}\{ 1-[1-D(p,t)]^K\} .
$$

Let us consider now a model with self-regulation. Moreover, each node also has
 $K$ links connected to  any of the $N$ nodes of the network. We focus on node $i$. This node has
probabilities $z_i=z$ and $w_i=w$ to belong to sets $\mathcal{E}(t)$ and $\mathcal{F}(t)$, respectively.
We want again to compute $W$.
If node $i$ is in set $\mathcal{F}(t)$ it has probability $1/2$ to contribute to $W$, independently of the $K$ links. Otherwise, at
least one of the $K$ links must be in set $\mathcal{F}(t)$. Taken in account these two situations, we have that
$$
W=\frac{w_i}{2}+\frac{z_i}{2}\left[{K \choose 1}z^{K-1}w+{K \choose 2}z^{K-2}w^2+\ldots + w^K\right].
$$
Identifying $W$ with $D(p,t+1)$ and $w$ with $D(p,t)$, we find that the Hamming distance
is given by
\begin{equation}
D(p,t+1)=\frac{1}{2}\{ 1-[1-D(p,t)]^{K+1}\} .\label{eq40}
\end{equation}
Observe that again this expression is similar to the previous one if $K$ is replaced by $K+1$.

Both models can be studied in a  scale-free topology. It is easy to obtain that
\begin{equation}
D(p,t+1)=\frac{1}{2}\left\{ 1-\sum_kP(k)[1-D(p,t)]^\theta \right\} ,\label{eq41}
\end{equation}
where $\theta =k+1$ for the case with self-regulation, and $\theta =k$ otherwise.
 These results are easily generalized to taken in account  the Derrida parameter
$p_d$ (see Derrida and Pomeau \cite{derr}).
In this case the probability $1/2$  of Eqs. (\ref{eq40}) and (\ref{eq41}) must be replaced by the
corresponding probability $2p_d(1-p_d)$.

\section{Mean-field  and simulation results}

\subsection{Results for $k_{min}=2$}

The fixed points for the model without a scale-free topology are obtained by putting
$M(p,t+1)=M(p,t)=M_*$ in the map given by Eq. (\ref{eq13}). 
 The fixed point $M_*=0$ always exists, and the local stability parameter $\lambda$, given by
$$
\lambda=\left.\frac{dM(p,t+1)}{dM(p,t)}\right|_{M_*}=pK(1-2M_*)^{K-1}+Kq {M_*}^{K-1},
$$
tell us that $M_*=0$ is stable for $\lambda=pK<1$. It means that $\lim_{t\to\infty} M(p,t)=0$ for any
initial value $M(p,0)$. 
 When $pK>1$ the initial conditions are attracted to a non null fixed point. 
 It means that in the $(p,K)$ plane, there is a curve, given by equation $pK=1$, separating the
region in which $M_*=0$ is stable  from the one that $M_*=0$ is not stable. These features can be
easily illustrated for $K=2$.
In this case, the non null fixed  point is given by $M_*=(1-2p)/(1-3p)$ and the local stability
parameter, evaluated for non null $M_*$, is $\lambda =2(1-p)$. Then we have that $\lambda < 1$ for $p>1/2$.
It implies that for $p>1/2$, the non null fixed point is attractive. For $K>2$ the evaluation of the fixed
point and of $\lambda$ were performed numerically.
\begin{table}
\caption{Values of density of $1's$ ($M(p)$) and Hamming distance ($D(p)$) for the model without a scale-free
topology with
$k_{min}=2$, $N=10000$, and the dynamics described by XOR and AND functions without self-regulation. The subscript {\it sim} is the simulated result and the {\it ann} refers to annealed (MF) solution. The error in the last digit of an evaluated quantity in the simulations is between parentheses.}
\label{tab1}
\begin{center}
\begin{tabular}{|c|c||c|c||c|c|}
\hline
$K$ & $p$ & $M_{ann}$ & $M_{sim}$ & $D_{ann}$ & $D_{sim}$ \\
\hline\hline 
$2$ & $0.0$   & 0.000      & 0.000(1)     & 0.000       & 0.000(2)    \\
  & $0.2$   & 0.000      & 0.000(1)     & 0.000       & 0.000(2)      \\
  & $0.5$   & 0.002      & 0.054(3)     & 0.002       & 0.060(3)       \\
  & $0.7$   & 0.364      & 0.362(3)     & 0.399       & 0.397(2)      \\
  & $1.0$   & 0.500      & 0.500(2)     & 0.500       & 0.500(1)     \\
\hline
$5$ & $0.0$   & 0.000      & 0.000(1)     & 0.000       & 0.000(1)  \\
  & $0.2$   & 0.000      & 0.017(3)     & 0.000       & 0.016(2)    \\
  & $0.5$   & 0.241      & 0.241(2)     & 0.242       & 0.241(2)     \\
  & $0.8$   & 0.402      & 0.402(2)     & 0.404       & 0.404(2)    \\
  & $1.0$   & 0.500      & 0.500(1)     & 0.500       & 0.500(2)    \\
\hline
$10$ & $0.0$  & 0.000      & 0.000(1)     & 0.000       & 0.000(1)  \\
   & $0.2$  & 0.084      & 0.084(1)     & 0.084       & 0.083(2)    \\
   & $0.5$  & 0.250      & 0.249(1)     & 0.250       & 0.250(2)     \\
   & $0.8$  & 0.400      & 0.400(1)     & 0.400       & 0.400(1)     \\
   & $1.0$  & 0.500      & 0.500(1)     & 0.500       & 0.500(1)     \\
\hline
\end{tabular}
\end{center}
\end{table}
In table (\ref{tab1}) we compare the results of the annealed approximation
for three values $K$ with the ones obtained by numerical simulations of
$N=10^4$ nodes, with the average quantities evaluated after $10^3$ mcs
in $3000$ samples. In the simulation results, the numbers between parentheses are the errors that affect the last digits.
 Similar results were obtained for other values of $p$.
 We can conclude that the MF results for the fraction of $1$'s  agree very well with those from the numerical simulations.

The fraction of $1$'s in a scale-free network is described by Eq. (\ref{eq15}).
Again the fixed point $M_*=0$ is always present. $\lambda$ can also evaluated and
we have that the $M_*=0$ is stable if $p<1/\langle k\rangle$. When $M_*=0$ is 
not stable, we obtain numerically that there is a non null fixed point attracting all the initial conditions.
These regions are separated in the $(p,\langle k\rangle)$ plane  by a curve described 
by
\begin{equation}
p\langle k\rangle =1.\label{eq43}
\end{equation}
In Tab. (\ref{tab2}) we compare the results of the annealed approximation with the
ones obtained from numerical simulations with $10^4$ nodes and $r=0.5$ ($\gamma=3$).
The $M_*$ results obtained by MF solution agree well with the ones from simulations.
 Similar results are obtained for other values of $r$.
\begin{table}
\caption{Values of density of $1's$ ($M(p)$) and Hamming distance ($D(p)$) for  Boolean dynamics
described by XOR and AND functions without self-regulation
on scale-free networks with $N=10000$, $k_{min}=2$, and $r=0.5$ ($\gamma=3$). 
The subscripts {\it sim} and {\it ann} refer to the simulated and MF results, respectively.
The errors evaluated in the simulations are between parentheses.}
\label{tab2}
\begin{center}
\begin{tabular}{|c||c|c||c|c|}
\hline
$p$ & $M_{ann}$ & $M_{sim}$ & $D_{ann}$ & $D_{sim}$ \\
\hline\hline
$0.2$  &  0.003          &  0.038(3)        &   0.038           &   0.007(3)          \\
$0.3$  &  0.075          &  0.085(3)        &   0.113           &   0.039(3)          \\
$0.5$  &  0.227          &  0.216(3)        &   0.242           &   0.178(3)          \\
$0.7$  &  0.352          &  0.350(4)        &   0.353           &   0.340(4)        \\
$0.8$  &  0.406          &  0.409(4)        &   0.404           &   0.414(4)         \\
\hline
\end{tabular}
\end{center}
\end{table}

The Eqs. (\ref{eq25}) and (\ref{eq27}) for the Hamming distance can be numerically solved
to furnish the fixed points in the cases of the models without and with scale-free
topology. Note that we are using that $w_1=w_0$ in both cases.
In Tabs. (\ref{tab1}) and (\ref{tab2}) we can compare the results obtained from
the annealed approximation with those obtained by the numerical simulations. We see that
they agree well. This conclusion holds for other values of the parameter $p$.

\subsection{Results for $k_{min}=1$}

In this case we have self-regulation. The fixed points for $M_*$ and $D_*$, obtained from
Eqs. (\ref{eq30}) and (\ref{eqnew1}) with $P(k)=\delta_{k,K}$, are displayed in Tab. (\ref{tab3}).
 We can see that MF results are different from the ones
obtained from numerical simulations for small $K$ ($K=1$ and $K=3$), although they are
similar when $K$ is large ($K=10$). In order to check our analytical approximation, we
have also performed numerical simulations with an annealed dynamics. At each mcs we
have randomly chosen the $K$ nodes connected with each node of the network. As we can see,
the numerical results agree very well with the ones obtained from MF.
\begin{table}
\caption{Values of density of $1's$ ($M(p)$) and Hamming distance ($D(p)$) for the model without a scale-free topology with 
$k_{min}=1$, $N=10000$, and the dynamics described by XOR and AND functions with self-regulation. The subscript {\it sim} is the simulated result and the {\it ann} refers to annealed (MF) solution.}
\label{tab3}
\begin{center}
\begin{tabular}{|c|c||c|c||c|c|}
\hline
$K$ & $p$ & $M_{ann}$ & $M_{sim}$ & $D_{ann}$ & $D_{sim}$ \\
\hline\hline
$1$ & $0.0$   & 0.000      & 0.002(2)     & 0.000       & 0.000(3)      \\
    & $0.2$   & 0.000      & 0.130(3)     & 0.000       & 0.058(3)       \\
    & $0.5$   & 0.002      & 0.277(4)     & 0.002       & 0.114(4)        \\
    & $0.8$   & 0.429      & 0.406(5)     & 0.454       & 0.172(5)        \\
    & $1.0$   & 0.500      & 0.497(4)     & 0.500       & 0.382(5)         \\
\hline
$3$ & $0.0$   & 0.000      & 0.000(2)     & 0.000       & 0.000(3)  \\
    & $0.2$   & 0.000      & 0.100(4)     & 0.000       & 0.073(4)  \\
    & $0.5$   & 0.230      & 0.250(3)     & 0.232       & 0.233(3)    \\
    & $0.8$   & 0.405      & 0.400(3)     & 0.410       & 0.399(3)     \\
    & $1.0$   & 0.500      & 0.500(3)     & 0.500       & 0.500(3)       \\
\hline
$10$ & $0.0$  & 0.000      & 0.000(2)     & 0.000       & 0.000(3)  \\
   & $0.2$  & 0.088      & 0.010(2)     & 0.088       & 0.099(3)   \\
   & $0.5$  & 0.250      & 0.250(3)     & 0.250       & 0.250(2)    \\
   & $0.8$  & 0.400      & 0.400(3)     & 0.400       & 0.400(3)     \\
   & $1.0$  & 0.500      & 0.500(3)     & 0.500       & 0.500(3)      \\
\hline
\end{tabular}
\end{center}
\end{table}

The same conclusions hold for the scale-free topology.
By using Eq. (\ref{eq30}) in the analysis of stability of the $M_*=0$ fixed point,
 we obtain that the curve separating the two regions is given by
\begin{equation}
p(\langle k\rangle +1) =1.\label{eq44}
\end{equation}
Even if we take into account that $\langle k\rangle$ of the above equation
begins with $k=1$ and the one of Eq. (\ref{eq43}) begins with $k=2$, Eqs.
 (\ref{eq43}) and (\ref{eq44}) are different. This implies that self-regulation
 changes the dynamical behavior. 

Table \ref{tab4} shows the results of computational simulation and the annealed approximation. It 
can be seen that the values of $M$ and $D$ obtained via MF when $k_{min}=1$ are a bit smaller than
the ones of the $k_{min}=2$ case. Another important feature is the relative good agreement between the values of $M$  obtained via 
simulation and that from annealed approximation. This match  does not occur for the Hamming distance. As we can see in
Tab. \ref{tab4}, the spreading damage calculated via MF is quite bigger than the  one obtained in 
simulations. We can conclude by analyzing the Hamming distance, that the self-regulated nodes introduce a new  dynamical behavior, 
with distinct properties when compared to the behavior of non-self-regulated ones. Is worth to comment that self-regulation is a common feature 
in biological networks. Maybe it is a process used in order to increase homeostasis, reducing the effect of a damage introduced in the system.
\begin{table}
\caption{Values of density of $1's$ ($M(p)$) and Hamming distance ($D(p)$) for  Boolean dynamics
described by XOR and AND functions with self-regulation
on scale-free networks with $N=10000$, $k_{min}=1$, and $r=0.5$ ($\gamma=3$). 
The subscripts {\it sim} and {\it ann} refer to the simulated and MF results, respectively.
The errors evaluated in the simulations are between parentheses.}
\label{tab4}
\begin{center}
\begin{tabular}{|c||c|c||c|c|}
\hline
 $p$ & $M_{ann}$ & $M_{sim}$ & $D_{ann}$ & $D_{sim}$ \\
\hline\hline
$0.2$  &  0.091          & 0.097(3)         &  0.090           & 0.000(3)          \\
$0.3$  &  0.141          & 0.140(2)         &  0.140           & 0.000(3)          \\
$0.5$  &  0.246          & 0.228(3)         &  0.243           & 0.001(3)          \\
$0.7$  &  0.351          & 0.325(2)         &  0.345           & 0.013(4)        \\
$0.8$  &  0.401          & 0.380(2)         &  0.396           & 0.042(4)         \\
\hline
\end{tabular}
\end{center}
\end{table}

\section{Summary}

In this work we studied  Boolean dynamics in Kauffman models and in scale-free networks.
 The dynamical models assigned only $XOR$ and $AND$ operators to the nodes  with probability $p$ and $1-p$.
 Regarding the inputs of the above cited Boolean networks, two types of dynamics were used. In the 
first one, the state of the nodes was regulated by the state of all nodes connected to them. The second type was 
similar to the first one, with the difference that the state of the node was used as its own input. Thus, in the first case we did not have self-regulation as in the second one. As shown in the results, these two types of dynamics presented quite different behaviors.
In both cases a computational simulation and an analytical mean-field approximation were performed in order to compare the density of $1's$, namely $M$, and the Hamming distance $D$. The results for the dynamics with no self-regulation generated good agreement between simulations and the MF approach. However, the case with self-regulation had a clear disagreement  with respect to $D$ and $M$, for small values of $p$.
\bigskip

\noindent The authors thank J. F. F. Mendes for useful discussions. We thank the referees for useful suggestions.
ACS and JKLS thank to Funda\c{c}\~ao de Amparo a Pesquisa de MG (FAPEMIG), a Brazilian agency, for partial financial support.
JKLS thanks to Conselho Nacional de Pesquisa (CNPq) for partial financial support.\\
\noindent$^{\ast}$Electronic address: alcidescs@gmail.com\\
\noindent$^{\dagger}$Electronic address: jaff@fisica.ufmg.br\\

\end{document}